\documentclass{article}

\usepackage{spconf}

\usepackage{romannum}
\usepackage{times}
\usepackage{amsmath,dsfont}
\usepackage{amssymb}
\usepackage{epsfig,verbatim} 
\usepackage{setspace}
\usepackage{color}
\usepackage{xcolor}
\usepackage{cite}
\usepackage{epstopdf}
\usepackage{graphics}
\usepackage{accents}
\usepackage{acronym,flushend}
\usepackage[bookmarks,colorlinks]{hyperref}
\usepackage{booktabs}
\usepackage{mathtools}
\usepackage[ruled,linesnumbered,vlined]{algorithm2e}
\usepackage{enumitem}
\usepackage[table]{xcolor}
\usepackage{svg}
\usepackage{xparse}

 \usepackage[all=normal,paragraphs=tight,floats=normal,mathspacing=normal,wordspacing=tight,charwidths=tight,mathdisplays=normal,leading=normal]{savetrees}

\newcommand{\myVec}[1]{{\boldsymbol{#1}}}
\newcommand{\myMat}[1]{{\boldsymbol{#1}}}
\newcommand{\mySet}[1]{\mathcal{#1}}
\newcommand{\Tsig}{{\tau_{\max}}}
\newcommand{\TDAmb}[1]{\myVec{a}_{#1}}
\newcommand{\TDAmbMat}[1]{\myMat{A}^{\tau}_{#1}}

\newcommand{\FDAmb}[1]{\myVec{A}^f_{#1}}
\newcommand{\FDAmbest}[1]{\hat{\myVec{A}}^f_{#1}}

\definecolor{NewColor}{rgb}{0,0,0}

\acrodef{kf}[KF]{Kalman filter} 
\acrodef{doa}[DOAs]{directions of arrival} 
\acrodef{sh}[SH]{spherical harmonics} 
\acrodef{sgm}[SGM]{score-based generative model} 
\acrodef{sde}[SDE]{stochastic differential equation} 
\acrodef{au}[AU]{Ambisonics upscaling} 
\acrodef{foa}[FOA]{first-order Ambisonics} 
\acrodef{hoa}[HOA]{high-order Ambisonics} 
\acrodef{stft-sdr}[STFT-SDR]{STFT signal-to-distortion ratio} 
\acrodef{stft-si-sdr}[STFT-SI-SDR]{STFT scale invariant signal-to-distortion ratio}

\acrodef{ve}[VE]{Variance Exploding} 
\acrodef{tf}[TF]{time-frequency}
\acrodef{stft}[STFT]{short-time Fourier transform}
\acrodef{istft}[ISTFT]{inverse STFT}
\acrodef{ncsn}[NCSN++]{noise-conditioned score-matching network}
\acrodef{rd}[RD]{reverse diffusion}
\acrodef{ald}[ALD]{annealed Langevin dynamics}
\acrodef{mushra}[MUSHRA]{MUltiple Stimuli with Hidden Reference and Anchor}
\acrodef{pwd}[PWD]{plane wave decomposition}
\acrodef{cs}[CS]{compressed sensing}

\acrodef{ekf}[EKF]{extended \ac{kf}}
\acrodef{ai}[AI]{artificial intelligence} 
\acrodef{dnn}[DNN]{deep neural network} 
\acrodef{nn}[NN]{neural network} 
\acrodef{rnn}[RNN]{recurrent neural network} 
\acrodef{mse}[MSE]{mean-squared error}
\acrodef{ista}[ISTA]{iterative soft-thresholding algorithm}
\acrodef{snr}[SNR]{signal-to-noise ratio}
\acrodef{ss}[SS]{state-space}
\acrodef{ml}[ML]{machine learning}
\acrodef{lufs}[loudness units relative to full scale]{LUFS}

\begin{document}
\title{DiffAU: Diffusion-Based Ambisonics Upscaling \vspace{-2mm}
}

\name{Amit Milstein, Nir Shlezinger, and Boaz Rafaely \vspace{-5mm}
}
\address{\small ECE School Ben-Gurion University of the Negev, Be’er-Sheva, Israel (e-mail: amitmils@post.bgu.ac.il; \{nirshl; br\}@bgu.ac.il).
}

\maketitle


\begin{abstract}
Spatial audio enhances immersion by reproducing 3D sound fields, with Ambisonics offering a scalable format for this purpose. While \ac{foa} notably facilitates hardware-efficient acquisition and storage of sound fields as compared to \ac{hoa}, its low spatial resolution limits realism, highlighting the need for \ac{au} as an approach for increasing the order of Ambisonics signals. 
In this work we propose {\em DiffAU}, a cascaded \ac{au} method that leverages recent developments in diffusion models combined with novel adaptation to spatial audio to generate 3rd order Ambisonics from \ac{foa}. By learning data distributions, DiffAU provides a principled approach that rapidly and reliably reproduces \ac{hoa} in various settings. 
Experiments in anechoic conditions  with multiple speakers, show strong objective and perceptual performance.
 \end{abstract}

\acresetall

\begin{keywords}
Spatial Audio, Ambisonics, Diffusion.
\end{keywords}

\vspace{-0.2cm}
\section{Introduction}

\label{sec:intro}
Spatial audio technology enhances the listener’s experience by accurately reproducing the direction and distance of sound sources in a three-dimensional space. It is commonly used in VR/AR, gaming, cinema, teleconferencing, and music to create immersive and realistic soundscapes~\cite{hacihabiboglu2017perceptual}. 
Among  spatial audio formats, Ambisonics \cite{zotter2019ambisonics} stands out for its flexibility and scalability in capturing, encoding, and rendering sound fields.
\Ac{foa}, which uses four channels, offers a practical advantage by requiring relatively simple hardware~\cite{gerzon1975design}. However, its spatial resolution is limited, leading to coarser localization and immersion.
In contrast, \ac{hoa} offers significantly better spatial detail through more channels~\cite{daniel2003further}. However, capturing \ac{hoa} requires large, expensive microphone arrays, limiting its accessibility.
This gap motivates the development of efficient upsampling techniques that can enhance \ac{foa}’s spatial resolution without the need for high-order acquisition hardware, enabling high-quality spatial audio at lower cost.

Several methods have been proposed in the literature for \ac{au}. Early approaches are predominantly model-based, and rely on physical assumptions on the sound field. A representative approach applies  \ac{cs} techniques for \ac{pwd} under the assumption that the sound field is sparse in the source domain \cite{wabnitz2012frequency, routray2020sparse, wabnitz2011upscaling}.  While \ac{cs} techniques enable upscaling under ideal conditions, such as free-field and low-noise settings, their performance deteriorates  when the sparsity assumption does not hold, limiting their applicability to real-world acoustic environments.

To address the limitations of model-based \ac{au}, several data-driven methods have been introduced in recent years. Gao et al.~\cite{gao2022sparse} proposed a multi-scale convolutional network operating in the frequency domain, incorporating sparse encoding to enhance generalization. While this method demonstrated improved performance over classical counterparts when the sparsity assumption holds, it similarly struggles in scenarios where this assumption fails. Routray et al.~\cite{routray2019deep} designed a multi-stage \ac{dnn} where each stage incrementally upsamples  by one order using fully connected networks. Despite its novel hierarchical structure, the architecture lacks expressivity and a theoretical basis. More recently, Nawfal et al.~\cite{nawfalambisonics} employed a waveform-domain encoder-decoder architecture adapted from Conv-TasNet~\cite{luo2019conv}, enabling upscaling in a latent space. Although showing significant improvement over \ac{foa}, the reported listening test showed some gaps from ideal 3rd order Ambisonics. These limitations motivate the search for methods that can further improve the quality of \ac{au}. Generative networks, which were not explored for this task to-date, offer a promising solution.

In this work, we propose {\em DiffAU}, a novel cascaded framework for \ac{au} that integrates the hierarchical structure of Ambisonics orders with the generative capabilities of diffusion models. By treating \ac{au} as a structured generative task, DiffAU enables principled upscaling from \ac{foa} to \ac{hoa} (3$^{rd}$ order in this work), through a sequence of intermediate stages, each implemented as a conditional diffusion process. This approach offers key conceptual advantages: it avoids explicit prior assumptions such as source sparsity and provides a probabilistic  mechanism for resolving the underdetermined nature of \ac{au}. Moreover, the modularity of DiffAU allows flexible extension to arbitrary upscaling ranges and facilitates order-by-order interpretability and training.

Our design begins by formulating the \ac{au} problem as  conditional sampling using \acp{sde}, where the goal is to sample \ac{hoa} coefficients conditioned on \ac{foa}. Based on this formulation, we develop a tailored diffusion model for spatial audio, incorporating signal representations aligned with Ambisonics encoding and appropriate transformations. Each diffusion stage is trained independently using denoising score matching, and the full system is realized via a cascaded architecture. Extensive experiments demonstrate that DiffAU systematically outperforms available \ac{au} baselines, 
highlighting the potential of generative diffusion-based techniques for spatial audio applications.

The rest of this paper is organized as follows: Section~\ref{sec:System Model and Preliminaries} introduces the signal model 
and some preliminaries. 
Section~\ref{sec: Method} describes  DiffAU, while Sections~\ref{sec: Numerical Study}-\ref{sec:listening_exp}  present its empirical study and listening test.  Section~\ref{sec: Conclusions} provides concluding remarks.

\vspace{-2mm}
\section{System Model and Preliminaries}
\label{sec:System Model and Preliminaries}  

\subsection{Signal Model and Ambisonics}
\label{subsec: Signal Model}
Consider a sound field composed of $Q$ plane waves with \ac{doa} $(\theta_q ,\phi_q)$, $q \in \{1,\ldots,Q\}$. We denote the source signals as $\mathbf{s}(k)$, a $Q$-dimensional vector 
where each element corresponds to the complex amplitude of a plane wave at the origin, with $k$ denoting the wave number. The Ambisonics signal of order $N$ due to $\myVec{s}(k)$ and the $Q$ plane waves can be written as~\cite{rafaely2015fundamentals}
\begin{equation}\label{eq:ambi_signal}
    \myVec{a}_{N}(k) = \myMat{Y}_Q^H\myVec{s}(k),
\end{equation} 
where $\myVec{a}_{N}(k)$ is of size $(N + 1)^2$, $\myMat{Y}_Q \in \mathbb{R}^{Q\times(N+1)^2}$ is a matrix composed of spherical harmonic functions evaluated at the microphone directions $(\theta_q,\phi_q)$ \cite{rafaely2015fundamentals}, and $(\cdot)^H$ denotes the conjugate transpose.. This Ambisonics signal can be computed from microphone signals using a spherical array of radius $r$~\cite{rafaely2015fundamentals}, and to avoid spatial aliasing for $k\cdot r\ll N$, it must hold that $(N + 1)^2 \leq M$, where $M$ is the number of microphones in the array.

\vspace{-0.1cm}
\subsection{Problem Formulation}
\label{ssec:Problem}
\ac{au} refers to the mapping of a low-order Ambisonics signal of order $N$ into higher-order coefficients to obtain a signal of order $N' > N$. This as is formulated an inverse problem
\begin{equation}\label{eq:inverse_prob_formulation}
\myVec{a}_{N}(k) = \myMat{F}\myVec{a}_{N'}(k),
\end{equation}
where $\myVec{a}_{N}(k)$ is an Ambisonics signal of order $N$ with $(N+1)^2$ channels, $\myVec{a}_{N'}(k)$ is an Ambisonics signal of order $N'$ with $(N'+1)^2$ channels, and $\myMat{F} \in \mathbb{R}^{(N+1)^2\times(N'+1)^2}$ is a sampling matrix which takes the first $(N+1)^2$ channels of $\myVec{a}_{N'}(k)$.

Since $\myMat{F}$ is a wide matrix, the problem is underdetermined. 
A straightforward method for tackling it is the least-norm approach; however, as demonstrated in \cite{epain2009application}, it tends to distribute the energy uniformly across the plane-wave sources, resulting in distortions. 
Therefore, to obtain a physically plausible solution, one should incorporate prior knowledge. To motivate our approach for learning the  distribution
$p(\myVec{a}_{N'}(k)\mid \myVec{a}_{N}(k))$, introduced in Section~\ref{sec: Method},
we next review \acp{sgm}.

\vspace{-0.3cm}
\subsection{Preliminaries of \acp{sgm}}\label{ssec:sgm_model}
\acp{sgm} \cite{song2019generative} are diffusion-based generative models that learn to reverse a noise corruption process. The formulation of \cite{song2020score_sde} casts this process into a continuous-time \ac{sde}, providing a unifying framework for SGMs. The forward process is expressed as
\begin{equation}\label{eq:forward_sde}
d\myVec{x}_t = f(\myVec{x}_t,t)dt + g(t)d\myVec{w},
\end{equation}
where $t$ denotes the continuous time index, $f$ is the drift term, $g$ the diffusion coefficient, and $\myVec{w}$ a  Wiener process.
The  reverse-time dynamics \cite{anderson1982reverse} are
\begin{equation}\label{eq:reverse_sde}
d\myVec{x}_t = [f(\myVec{x}_t,t) - g(t)^2\nabla_{\mathbf{x}} \log p_{t}(\mathbf{x_t})]dt +  g(t)d\myVec{\bar{w}},
\end{equation}
with $\myVec{\bar{w}}$ a time-reversed Wiener process, $p_t(\mathbf{x}_t)$ the marginal probability density of $\mathbf{x}_t$ at time $t$, and $\nabla_{\mathbf{x}}$ the gradient operator.
The score function $\nabla_{\myVec{x}} \log p_{t}(\myVec{x}_t)$ is approximated by a neural network $\myVec{s}(\myVec{x}_t,t;\myVec{\theta})$. Sampling can then be carried out using Predictor-Corrector samplers \cite[Appendix G]{song2020score_sde}.

\vspace{-0.1cm}
\section{Proposed Method}
\vspace{-0.2cm}

\label{sec: Method}
\acp{sgm} have been  proposed for image super-resolution, e.g., \cite{ho2022cascaded,moser2024diffusion}. However, applying this methodology to  \ac{au} is not straightforward, and existing image super-resolution methods do not directly transfer to audio. Still, the success of this approach in other domains motivates exploring its adaptation to spatial audio.
To that end, we introduce the proposed DiffAU framework.

\vspace{-0.3cm}
\subsection{DiffAU}

\label{ssec:diffau}
To explain how we adapt diffusion-based super resolution to  spatial audio, we begin by formulating the diffusion model and \ac{sde}. Next, we describe the data representation employed in our diffusion models, and  present our proposed DiffAU for \ac{au}. 

\begin{figure}
    \centering
    \includegraphics[width=\linewidth]{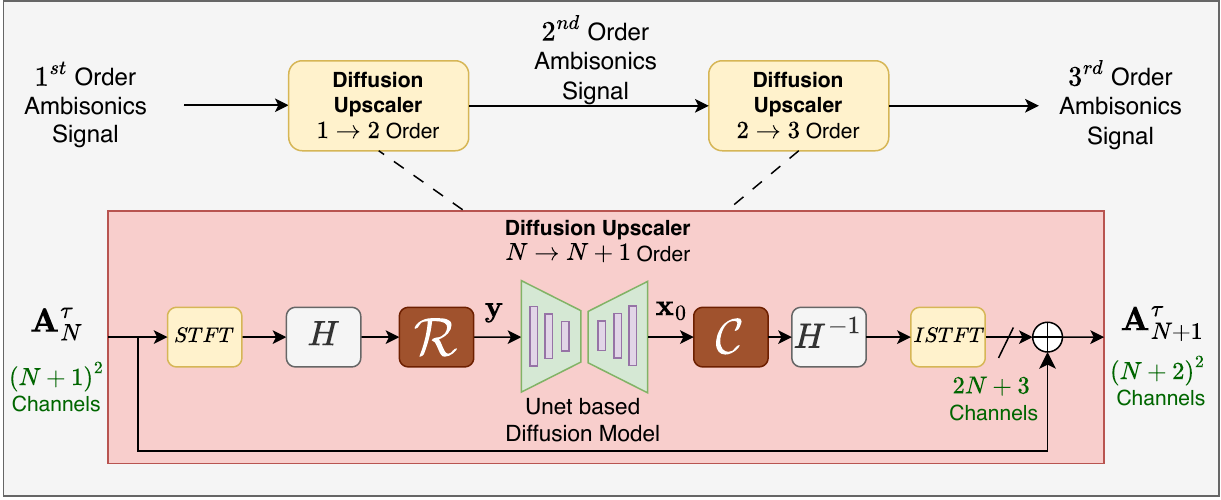}
    \caption{Schematic illustration of the overall architecture of DiffAU}
    \vspace{0.3cm}
    \label{fig:DiffAU}
\end{figure}

{\bf \ac{sde} Formulation for Spatial Audio}:
The formulation presented in Subsection~\ref{ssec:sgm_model} is geared towards sampling from a prior distribution of some variable $\myVec{x}$, which in our case represents higher-order Ambisonics. However,  we are interested in sampling from the posterior distribution $p(\myVec{x}|\myVec{y})$ conditioned on an observation $\myVec{y}$, e.g., a lower-order Ambisonics. This setting requires a \textit{conditional diffusion model} which must estimate the gradient $\nabla_{\myVec{x}} \log p_{t}(\myVec{x}|\myVec{y})$. Conditional diffusion can be achieved by concatenating the observation 
$\myVec{y}$ to the score model input \cite{saharia2022image}.  In our proposed method, we adopt this strategy by using a lower order Ambisonics, \ac{foa} in our case, as the observation.

The \ac{sde} we use is the \ac{ve} \ac{sde} introduced in\cite{song2020score_sde}. In this formulation, the drift and diffusion terms in the forward \eqref{eq:forward_sde}  and reverse  \eqref{eq:reverse_sde} SDEs are defined as follows 
\begin{align}
f(\myVec{x}_t,t)= 0, \quad
g(t) = \sigma_{\text{min}} \left( \frac{\sigma_{\text{max}}}{\sigma_{\text{min}}} \right)^t \sqrt{2 \log \left( \frac{\sigma_{\text{max}}}{\sigma_{\text{min}}} \right)}. \label{eq:diffau_diff_coeff}
\end{align} 
Here, $\sigma_{\min}$ and $\sigma_{\max}$ denote the minimum and maximum noise levels specified by the scheduler.

\textbf{Data Representation}: The input signals are order-$N$ real Ambisonics with N3D normalization~\cite{zotter2019ambisonics}, represented in the time domain as a matrix $\TDAmbMat{N} \in \mathbb{R}^{(N+1)^2 \times \Tsig}$, where $\Tsig$ denotes the total number of time samples. We apply the \ac{stft} to obtain the frequency-domain representation $\FDAmb{N}$, followed by a nonlinear amplitude transformation $\mathcal{H}(x)$~\cite{richter2023speech,dong2025edsep} to normalize amplitudes and ensure consistent DNN input scales:
\begin{align}
\mathcal{H}(x) = \frac{|x|^\alpha}{\beta} e^{j \angle(x)}, \quad
\mathcal{H}^{-1}(x) = \beta |x|^{1/\alpha} e^{j \angle(x)}\label{eq:bwd_transformation}
\end{align}
where $\alpha$ and $\beta$ are hyperparameters and $\angle(\cdot)$ denotes the phase operator. Additionally, we define the operator $\mySet{R}$ to convert complex tensors to real-valued inputs for the \ac{dnn} by concatenating real and imaginary parts along the channel dimension, and $\mySet{C}$ as its inverse.

\textbf{Overall Algorithm}: DiffAU, illustrated in Fig.~\ref{fig:DiffAU}, consists of two cascaded diffusion blocks, each upscaling the Ambisonics order by one ($N\mapsto N+1$) to produce a third-order representation from a \ac{foa} input.
Within each block, a score model $\myVec{s}_N(\;\cdot\;,\myVec{y};\theta_N)$—based on the \ac{ncsn} architecture \cite{song2020score_sde}—is conditioned on the amplitude-transformed current-order signal ($\myVec{y}$) to predict the additional $(2N+3)$ channels needed for the next order.

The generation and cascaded composition process is outlined in Algorithm~\ref{alg:diffau}. For the sampling within each block, we adopt the predictor-corrector framework \cite[Appendix G]{song2020score_sde}, employing reverse diffusion as the predictor and annealed Langevin dynamics as the corrector. In our implementation, we perform 30 predictor steps per diffusion block, with one corrector step per predictor iteration.

\begin{algorithm}
\caption{DiffAU}
\label{alg:diffau} 

\SetKwInOut{Input}{Input}
\Input{\ac{foa} $\TDAmbMat{1} = [\TDAmb{1}(1),\ldots,\TDAmb{1}(\Tsig)] \in \mathbb{C}^{4\times \Tsig}$\\
Score models 
$\myVec{s}_1(\;\cdot\;;\myVec{\theta}_1)$, $\myVec{s}_2(\;\cdot\;;\myVec{\theta}_2)$ 
}
\For{$N = 1$ \KwTo $2$}{%
 $\FDAmb{N} \leftarrow {\rm STFT}(\TDAmbMat{N})$ \;
 $\myVec{y} \leftarrow \mySet{R}\{\mySet{H}(\FDAmb{N})\}$\; 
 $\myVec{x}_T \in \mathbb{R}^{2(2N+3)} \sim \mathcal{N}(\myVec{0},\sigma_{\text{max}} \myVec{I})$\;
 $\myVec{x}_0 = \text{Sample}(\myVec{x}_T, \myVec{s}_N(\;\cdot\;,\;\cdot\;,\myVec{y};\myVec{\theta}_N))$ \;
 $\FDAmb{N+1} \leftarrow {\rm Concat}(\FDAmb{N}, \mySet{H}^{-1}(\mySet{C}(\myVec{x}_0 )))$\;
 $\TDAmbMat{N+1} \leftarrow {\rm ISTFT}(\FDAmb{N+1})$ 

}
\KwRet{\normalfont{\ac{hoa}} $\TDAmbMat{3} \in \mathbb{C}^{16\times \Tsig}$}
\end{algorithm}

\vspace{-0.1cm}
\subsection{Training}
\label{subsec: Training}
\vspace{-0.1cm}

DiffAU's \acp{sgm} $\{\myVec{\theta}_N\}_{N=1}^2$ are trained independently via denoising score matching \cite{vincent2011connection}, leveraging the Gaussian noise structure to sample $\myVec{x}_t$ in a single step \cite{shaked2025ai}. To upscale to order-$N+1$, we construct a dataset $\mathcal{D} = \{(\myVec{x}_0^{(i)},\myVec{y}^{(i)})\}_{i=1}^D$, where $\myVec{y}$ is the order-$N$ signal and $\myVec{x}_0 \in \mathbb{R}^{2N+3}$ represents the additional HOA channels. The empirical risk minimized over $\mathcal{D}$ is:
\begin{equation}
\hspace{-0.2cm}
    \mathcal{L}_{\mySet{D}}(\myVec{\theta}_N) \!=\! \frac{1}{D}\sum_{i=1}^{D}\left\|  s_N\big(\myVec{x}_{t^{(i)}}^{(i)},t^{(i)},\myVec{y}^{(i)}; \myVec{\theta}_N)\cdot\sigma_{t^{^{(i)}}} \!+\! \myVec{z}^{(i)} \right\|^2,
\end{equation}
where $t^{(i)} \sim \mathcal{U}[1,T]$ and $\myVec{z}^{(i)} \sim \mathcal{N}(0,\myMat{I})$, with $\mathcal{U}$ and $\mathcal{N}$ denoting the uniform and multivariate 
normal distributions, respectively


 \subsection{Discussion}
 \label{subsec: Discussion}
\vspace{-0.1cm}
DiffAU leverages the generative capabilities of diffusion models to sample from the posterior distribution, enabling physically plausible solutions to the ill-posed \ac{au}. It adapts diffusion models   to match the Ambisonics signal format and employs a cascaded structure to  estimated the \ac{hoa} channels. The cascaded architecture introduces modularity, allowing upscaling from any desired order. 
While we consider noiseless, multiple-speaker scenarios in free-field conditions, we expect its design to yield accurate \ac{au} also under noise and reverberation, while leaving this extension for future work. 

\vspace{-0.2cm}
\section{Numerical Study}
\label{sec: Numerical Study}
\vspace{-0.2cm}
We evaluate the proposed DiffAU in a numerical study\footnote{The source code and the complete set of hyperparameters used in our study is available at \url{https://github.com/Amitmils/GenAU}.} detailed next, and in a listening experiment detailed in Section~\ref{sec:listening_exp}.

{\bf Data}:  We constructed a dataset based on the WSJ0 corpus \cite{garofolo1993csr}. The dataset is split into training, validation, and test sets, with each speaker appearing in only one set and contributes multiple utterances to it. To generate the Ambisonic signal sets, we randomly selected $1-4$ speakers for each signal. For each selected speaker, a random DOA was assigned, and the corresponding 3rd order Ambisonics was constructed via \eqref{eq:ambi_signal}. This study focuses exclusively on free-field scenarios. All audio signals are 2.048 seconds long and sampled at 16 kHz.

{\bf Evaluation}:  Our evaluation uses the \ac{foa} (first four channels) as input. The model then estimates the remaining twelve channels  to reconstruct the \ac{hoa}. These predicted channels are compared to the corresponding ground truth using the \ac{stft-sdr} metric which measures the reconstruction quality, with higher values indicate a more faithful recovery. We compute it over all higher-order channels (i.e., channels 5–16) for each  sample:
\begin{equation}
\text{STFT-SDR}(\hat{\myMat{A}}^f_{3}) \!=\! 10 \cdot \log_{10} \left( 
\frac{\|  \FDAmb{3_{(5:16)}}\|_F^2}{\|\FDAmb{3_{(5:16)}} \!-\! \FDAmbest{3_{(5:16)}} \|_F^2}
\right)
\end{equation}
where $\FDAmb{3}$ and $\FDAmbest{3_{(5:16)}}$ are the \ac{tf} true and estimated Ambisonic signals, respectively. The subscript ${(5:16)}$ indicates the  channels used, and $\| \cdot\|_F$ is the Frobenius norm.

{\bf Results}:  
The \ac{stft-sdr} results presented in Table \ref{tab:audio_analysis} are based on 500 audio samples, corresponding to 0.25 hours of  data, with evaluation focused on the \ac{hoa} channels (channels 5--16). DiffAU was trained with 10 hours of data per diffusion block. 

We compare its performance against two distinct methodologies: first, the \ac{pwd} method using \ac{cs} in the frequency domain \cite{wabnitz2012frequency}, which is a model-based iterative method that addresses the underdetermined nature of \ac{au} by imposing sparsity on the sound field. Second, we evaluate a Conv-TasNet architecture inspired by \cite{nawfalambisonics}. Unlike the model-based approach, this is a data-centric, fully deterministic deep learning method that operates without any prior sparsity assumptions. 

\begin{figure}
    \centering
    \includegraphics[width=1\linewidth]
    {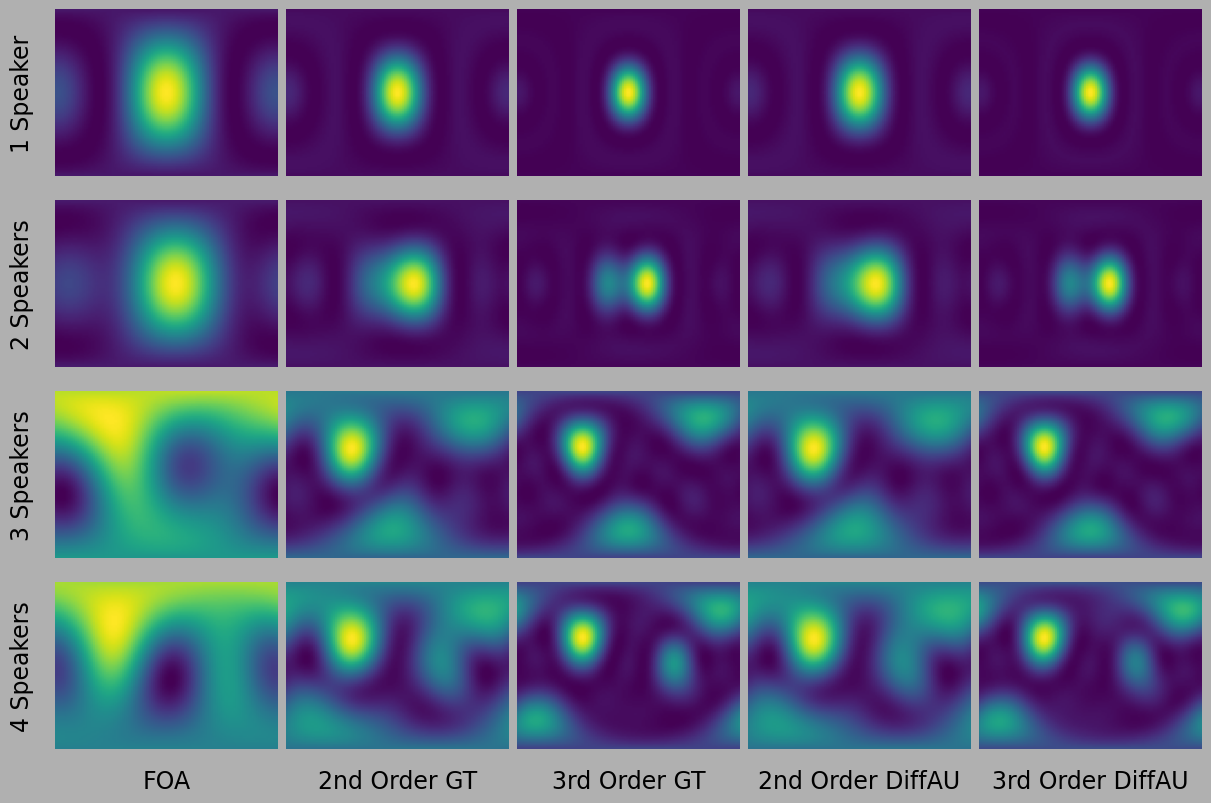}
    \caption{\small Directional energy plots (azimuth-elevation).}
    \label{fig:ambisonic_comparison}
\end{figure}

As shown in Table \ref{tab:audio_analysis}, DiffAU outperforms these baselines across all cases. Notably, even in a free-field setting where sparsity assumptions typically hold, DiffAU's ability to directly learn the posterior allows it to surpass both the classical \ac{pwd}-\ac{cs} and the deterministic Conv-TasNet. 
We also compare in  Fig.~\ref{fig:ambisonic_comparison} the directional energy plots of 2nd- and 3rd-order Ambisonics signals produced by DiffAU with the ground truth for orders 1–3, across one to four active speakers. The results show a strong resemblance between the recovered HOA and the reference energy patterns. 

\begin{table}[t]
\centering
\fontsize{7.5pt}{10pt}\selectfont
\begin{tabular}{@{}c c c c @{}}
\toprule
\textbf{\# Speakers} & \textbf{DiffAU} & \textbf{Conv-TasNet AU} & \textbf{PWD CS} \\ 
\midrule
1 & $29.5 \pm 6.7$  & $23.1 \pm 7.2$ & $12.9 \pm 7.9$  \\ 
2 & $27.3 \pm 3.8$ & $14.7 \pm 6.4$ & $14.3 \pm 2.2$  \\ 
3 & $23.1 \pm 4.0$ & $10.6 \pm 4.6$ & $12.3 \pm 2.6$  \\ 
4 & $19.6 \pm 4.5$ & $8.5 \pm 4.3$ & $10.9 \pm 2.6 $ \\ 
\midrule
\textbf{Overall} & $\mathbf{24.7} \pm \mathbf{6.2}$ & $14.0 \pm 7.9 $ & $12.6 \pm 4.5$  \\ 
\bottomrule
\end{tabular} 
\caption{\small \ac{stft-sdr} (dB) on \ac{hoa} channels vs. number of speakers. }
\label{tab:audio_analysis}
\end{table}

\section{Listening Test}\label{sec:listening_exp}
\vspace{-0.1cm}

In addition to the quantitative results shown in the Numerical Study, we performed a formal listening test to evaluate subjective audio quality and ensure that DiffAU did not introduce subtle artifacts undetectable by objective error metrics.

Similar to the \ac{mushra} protocol~\cite{series2014method}, participants evaluated overall signal quality relative to a reference across three \ac{mushra} screens. Each screen presented four signals: the 8th-order Ambisonics reference, a \ac{foa} anchor, the proposed 3rd-order signal and the true 3rd-order signal. All signals were rendered binaurally using the least-squares method~\cite{Avni2013spatial} and loudness-equalized to –6 \ac{lufs}; no head-tracking was applied. Each screen featured a single speaker from the test set at a colatitude of $90^\circ$ and azimuths of $15^\circ$, $30^\circ$, and $-60^\circ$.
Eight participants with no known hearing impairments and prior experience in spatial listening tests took part. The test was conducted in a quiet environment and included training and familiarization phases. During training, participants were introduced to the equipment and \ac{mushra} scale, while familiarization allowed free listening to all stimuli. Participants then rated the perceptual quality of each signal on a 0–100 scale, where 100 indicates no audible difference from the reference.

\begin{figure}
    \centering
    \includegraphics[width=0.8\linewidth, height=4cm]{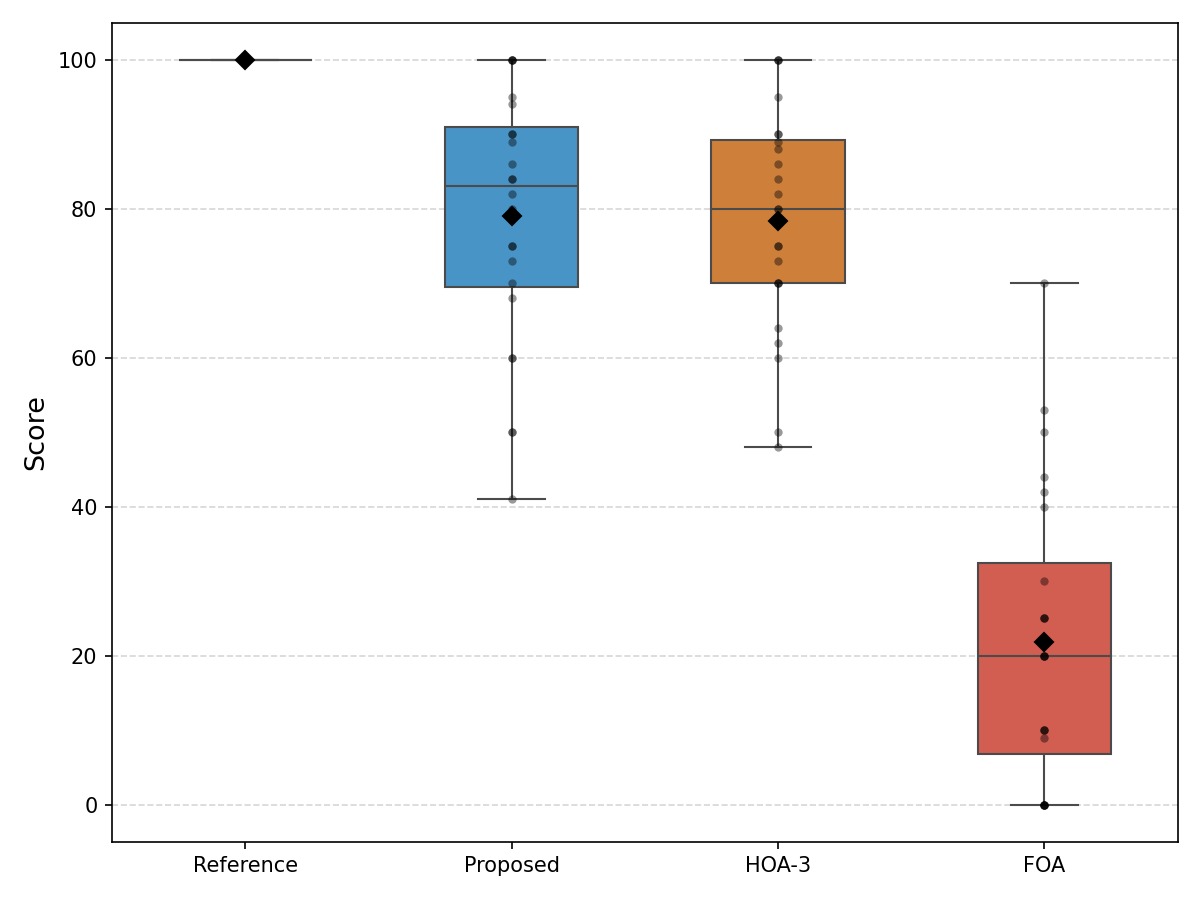}
    \caption{\small Listening test results} 
    \label{fig:mushra_results}
\end{figure}

Mauchly's test \cite{mauchly1940significance} indicated that sphericity was violated ($W = 0.095$, $p = 0.021$), so a Greenhouse-Geisser correction was applied\cite{greenhouse1959methods}. A repeated measures ANOVA revealed a significant effect of renderer on ratings ($F(3, 21) = 102.59$, $p < 0.001$). Post-hoc pairwise comparisons with Bonferroni correction showed that all pairs were significantly different ($p < 0.01$) except for the proposed method and the true 3rd-order signal ($p = 1.0$, Hedges' $g = -0.064$). As shown in Fig.\ref{fig:mushra_results}, the reference received a perfect score of 100, the proposed method and the true 3rd-order signal scored comparably (means of 79.0 and 78.4, respectively), while the 1st-order anchor averaged 21.8. However, it is acknowledged that this outcome may be influenced by the free-field conditions of the evaluation, and performance in more realistic acoustic environments remains to be investigated.

\section{Conclusion}
\label{sec: Conclusions}
We proposed a novel  \ac{au} method termed DiffAU. By leveraging  diffusion models to sample from the posterior distribution, DiffAU  addresses the inherent underdetermined nature of the \ac{au}  problem. For  multi-speaker scenarios in free-field conditions, DiffAU  outperforms the baselines, and  delivered  equivalent perceptual performance as the true \ac{hoa}. In future work, we plan to extend the proposed method to noisy and reverberant environments, aligning it more closely with real-world scenarios.


\bibliographystyle{IEEEtran}
\bibliography{refs}

\end{document}